\documentclass{article}
\pdfoutput=1

\usepackage[final]{neurips_2019}

\usepackage[bookmarks=false]{hyperref}

\usepackage[utf8]{inputenc} 
\usepackage[T1]{fontenc}    
\usepackage{hyperref}       
\usepackage{url}            
\usepackage{booktabs}       
\usepackage{amsfonts}       
\usepackage{nicefrac}       
\usepackage{microtype}      
\usepackage{amsmath}
\usepackage{graphicx}
\usepackage{caption}
\usepackage{float}
\usepackage{multirow}
\usepackage[export]{adjustbox}
\restylefloat{table}

\title{Conditional Denoising of Remote Sensing Imagery Using Cycle-Consistent Deep Generative Models}

\author{%
  Michael Zotov\thanks{Work completed as part of the Cervest Research Residency program. Correspondence to \texttt{jev@cervest.earth}} \qquad Jevgenij Gamper\thanks{Mentor} \\
  Cervest Ltd.\\ 
  London, UK \\
}

\begin{document}

\maketitle

\begin{abstract} 
The potential of using remote sensing imagery for environmental modelling and for providing real time support to humanitarian operations such as hurricane relief efforts is well established. These applications are substantially affected by missing data due to non-structural noise such as clouds, shadows and other atmospheric effects. In this work we probe the potential of applying a cycle-consistent latent variable deep generative model (DGM) for denoising cloudy Sentinel-2 observations conditioned on the information in cloud penetrating bands. We adapt the recently proposed Fr\'{e}chet Distance metric to remote sensing images for evaluating performance of the generator, demonstrate the potential of DGMs for conditional denoising, and discuss future directions as well as the limitations of DGMs in Earth science and humanitarian applications.
\end{abstract}

\section{Introduction}
Climate variability induced impacts are becoming increasingly important for land resource management for commercial aims and social good.  One mechanism for greater insights is employment of remote sensing imagery \cite{quarmby_use_1993, delecolle_remote_1992, ruswurm_early_nodate, doraiswamy_crop_2003, wang_crop_2019,malenovsky_sentinels_2012}. However, Earth observation data in high temporal and spatial resolution is not widely available. The effectiveness and accuracy of remote sensing based solutions (for instance, machine learning solutions for crop yield prediction) is hindered by the frequent appearance of atmospheric noise in such data, in the form of outliers induced by small shadows or missing data due to clouds. Besides the social benefit of efficient use of land resources through remote sensing, multiple humanitarian missions directly benefit from the support of satellite Earth observations \cite{rudner_multi3net:_2018}. Thus publicly available missions like Sentinel-2, MODIS, Landsat are of crucial importance, as these are accessible to non-profit or research organisations. However, it has been shown that even with  partially clouded skies there is just a 25\% chance that an arbitrary active field will appear in daily optical satellite imagery \cite{wal_fieldcopter:_2013}, which highlights the influence of cloud cover on the availability of optical imagery.

One approach that has been widely adopted is cloud masking. However, obtaining pixel-based segmentation ground truth in remote sensing is not as straightforward as for natural images. Current methods mostly resort to computationally intensive models that use temporal information or combine multiple physics-based models to create supervised learning data, analogous to the idea of data programming \cite{hagolle_multi-temporal_2010, noauthor_sen2cor_nodate, zupanc_improving_2019}. Furthermore, recent work has demonstrated that for the purpose of crop classification, deep learning models could deal with cloud induced outliers and still obtain sufficient performance \cite{ruswurm_multi-temporal_2018}. While these approaches prevent only some outliers in the downstream tasks, it does not solve the problem of missing data due to clouds, which fundamentally undermines the benefit of higher temporal resolution.

In this work, we therefore propose a novel treatment of surface reflectance data denoising from atmospheric artifacts as an unpaired translation from the domain of cloudy/shaded images $X$ to the domain of clear images $Y$. We contribute by: (i) demonstrating preliminary results of applying cycle-consistent deep generative models to this problem, (ii) building a deep convolutional neural network to obtain state of the art performance in land cover classification on the existing public dataset, and (iii) developing a method to assess the fidelity of the denoised remote sensing images by calculating the Fr\'{e}chet Distance between the distributions of generated and real images in the classifier feature space. 

\section{Methods and Results}

\subsection{Remote sensing data}
Sentinel-2 (S-2) is a European Space Agency program that provides surface reflectance data. Compared to other existing programs such as Landsat and MODIS, it provides 6-8 day temporal resolution at higher spatial resolution per pixel of 60m to 10m, depending on the band. S-2 samples data in 13 channels in the visible, near-infrared and short wave infrared spectral range. For this study, whilst the raw S-2 data is publicly available\footnote{https://registry.opendata.aws/sentinel-2/}, we chose a curated dataset created from S-2 observations, that has been established to standardise land cover classification performance in remote sensing - BigEarthNet \cite{sumbul_bigearthnet:_2019}. The dataset contains 590,326 images with multi-class labels of 43 ground cover categories and demonstrates the challenges of supervised learning for remote sensing classification tasks: some labels are present in over 200,000 examples while others appear in less than 1,000 examples. 

While there are other standardised public surface reflectance datasets, BigEarthNet also provides a breakdown of the dataset into images containing clouds and cloud shadows, and clear images\footnote{http://bigearth.net/} with respective cardinality 9,280 and 519,339 (images with seasonal snow are excluded) \cite{helber_eurosat:_2017}. This breakdown is beneficial to this problem, and represents the general challenge of cloud and shadow removal in remote sensing: it is possible to identify images that contain clouds or cloud shadows, but it is not easy to obtain pixel-wise ground truth to train segmentation models. 

\subsection{Land cover classification for denoising evaluation}
The performance of DGMs is difficult to measure \cite{salimans_improved_2016}, particularly in the context of remote sensing imagery, as an image that "appears" realistic to the human eye may still contain outliers for downstream tasks. Furthermore, given the social and economic importance of the task at hand, these models require a robust testing approach. As such we adopt the Fr\'{e}chet Distance (FD) based score, commonly used to evaluate DGMs, to remote sensing \cite{heusel_gans_2017}. FD captures similarity between the two curves. To evaluate DGMs using FD, one needs to train a sufficiently powerful image classifier, whose features would parametrise multivariate normal distributions of generated and real samples in the classifier feature space, and then compare the two distributions using FD. We therefore train an 18 layer ResNet \cite{he_deep_2015} on the BigEarthNet multi-label classification task, and achieve a state of the art performance. Instead of sigmoid activations in the final layer commonly used in multilabel classification, we use softmax activations \cite{joulin_learning_2015}. To regularise against dataset imbalance during training we randomly set images to only contain a single label with a probability of $0.5$, selected uniformly from the original list of labels for that image. Our results in land cover classification, compared to the model used by the creators of the dataset, are presented in Table 1.

\begin{table}[!htbp]
\label{resnet}
\centering
\caption{Experimental results of training the landcover classifier}
\begin{tabular}{| l | l | l | l | l |}
\hline
Model & $Precision(\%)$ & $Recall(\%)$ & $F_1$ & $F_2$    \\
\hline
S-CNN-All \cite{sumbul_bigearthnet:_2019} & $70$ & $77$ & $0.70$ & $0.74$  \\
\hline
Regularized ResNet (our model) & $85$ & $77$ & $0.79$ & $0.77$ \\
\hline
\end{tabular}
\end{table}

We then proceed to establish a baseline validation FD score for our task. We computed a FD score of 228 for cloudy versus clear validation sets.

\begin{figure}[!htbp]
    \label{training}
    \centering
    \begin{minipage}{0.45\textwidth}
        \centering
        \includegraphics[width=0.9\textwidth]{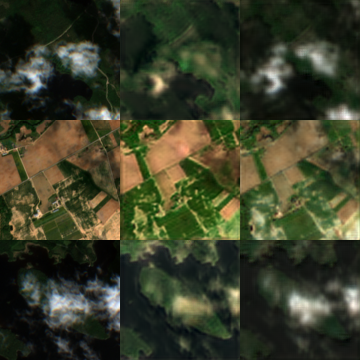}
        \vspace*{+1.2mm}
        \caption*{a)}
    \end{minipage}\hfill
    \begin{minipage}{0.45\textwidth}
        \hspace{-3em}
        \includegraphics[width=1.25\textwidth]{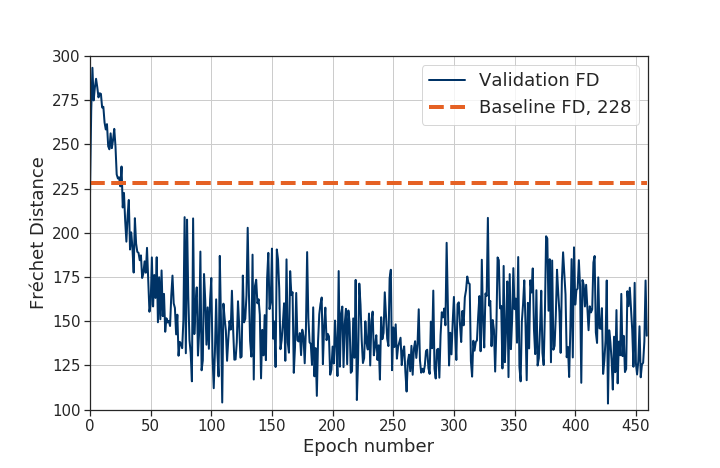} 
        \vspace*{+3.5mm}
        \caption*{b)}
    \end{minipage}
    \caption{a) The first column corresponds to training images in domain $X$, middle column is the same images mapped to domain $Y$, and in the third column these images are restored back to domain $X$ for the evaluation of cycle-consistency loss. Images were captured at the 100th epoch, with FD of 106. b) Fr\'{e}chet Distance over the course of training of an Aug CycleGAN.}
\end{figure}

\subsection{DGM for cloud and shadow removal}
We approach the problem from a novel perspective of unpaired image-to-image translation between the domain of cloudy/shaded images, $X$, and clear images $Y$. We acknowledge the randomness in spatial distribution of clouds or shadows by introducing a latent variable $\mathbf{z_x}$, and the distribution over possible land cover categories underneath the cloud as $\mathbf{z_y}$. As such, we have two joint distributions describing the respective pairs $(\mathbf{x}, \mathbf{z_y}) \sim p_{d}(\mathbf{x})p(\mathbf{z_y})$ of cloudy/shaded images and the possible land classes underneath, and $(\mathbf{y}, \mathbf{z_x}) \sim p_{d}(\mathbf{y})p(\mathbf{z_x})$ of clear images and possible locations of clouds and shadows respectively. Our task is to learn the implicit, parameterised generative models $p_{\theta}(\mathbf{x} | \mathbf{y}, \mathbf{z_x})$ and $p_{\phi}(\mathbf{y} | \mathbf{x}, \mathbf{z_y})$. By construction, every cloudy image in domain $X$, when mapped into the clear domain $Y$, is conditioned on the observed cloud penetrating bands. While in this preliminary work the model construction is not ideal, in future work we will include radar based bands that in fact do contain significant information about the underlying land cover and are not affected by clouds \cite{abdikan_land_2016}.

To infer the parameters of the implicit models we adopt the inference scheme of \textit{Augmented Cycle-Consistent Generative Adversarial Networks} (Aug CycleGAN) \cite{almahairi_augmented_2018}. Namely, we utilise 8 deep neural networks: two Generator - Discriminator pairs, where generators learn maps $X\times Z_y \rightarrow Y$ and $Y\times Z_x \rightarrow X$ respectively, and two Latent Encoder - Discriminator pairs. We add Spectral Normalization to the convolutional layers in the architectures \cite{miyato_spectral_2018}, and optimise the model using the loss functions described by the authors of the original paper. The model is optimised to map a sample between domains and then recover it consistently (see Figure 2a), where cycle-consistency is enforced by the L1 norm. During training, we iterate over a subset of cloudy images, and select a batch of clear images at random (from a training subset) for each batch of cloudy images. The validation set consists of fixed pairs, for better performance comparison between epochs. We compute FD for validation set after every epoch (see Figure 1b). 

\begin{figure}[!htbp]
\label{results}
    \centering
    \begin{minipage}{0.45\textwidth}
        \centering
        \includegraphics[width=0.9\textwidth]{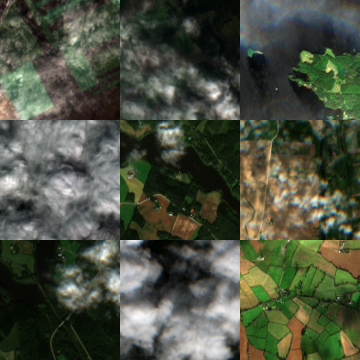} 
        \caption*{a)}
    \end{minipage}\hfill
    \begin{minipage}{0.45\textwidth}
        \centering
        \includegraphics[width=0.9\textwidth]{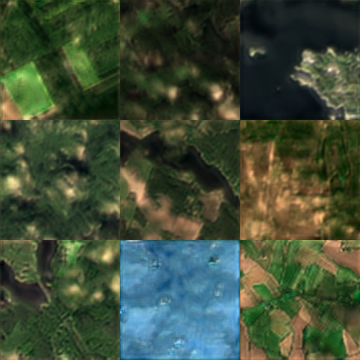} 
        \caption*{b)}
    \end{minipage}
    \caption{a) Red green and blue bands of validation images originally in the domain $X$ (cloudy and shaded). b) These same images mapped to domain $Y$ (clear).}
\end{figure}

In Figure 2b we show that the generator is able to capture information from cloud penetrating bands to inpaint the land segment covered by the cloud (e.g. images [1,3] (row 1 column 3) and [1,1]) as well as address shadows (image [3,3]), and preserve the landscape, but struggles with the thicker parts of the cloud, as might be expected (seen in [2,1] and [1,2]). The light brown spots in such areas are indicative of some mode collapse.
We also provide an example of a model failure to translate the image to the domain $Y$ ([3,2]).
\newpage
Additionally, we observed that lower FD does not always correspond to high fidelity samples in this particular problem. A model that has suffered mode collapse (characterised by inpainting the clouds by light brown) may have lower FD on validation than one that produces a diverse set of samples. Inception score, which measures the entropy of predicted classes over the generated images would be a better metric for this case \cite{salimans_improved_2016}. In future work we are aiming to extend analogous score to multi-label cases.

\section{Discussion \& Future Directions}
While we provide the evidence of success that the proposed approach addresses the problem of missing values and outliers in remote sensing data, it is important to point out the lack of precision in available tools to evaluate DGMs, particularly for critical applications of land modeling or humanitarian missions, or even for more sophisticated modelling applications in Earth sciences. Furthermore, the proposed approach is rooted in the assumptions of generative modeling, yet the capacity of DGMs to learn the underlying density has been put under question recently \cite{nalisnick_deep_2018, zhao_bias_2018}. Furthermore, due to the imbalance of land-cover classes in remotely sensed images and larger number of channels per image, issues of mode-collapse are only emphasized during adversarial training and addressing these simply by bigger models and larger scale training is not practical for remote sensing or Earth science applications \cite{brock_large_2018}. Resolving these issues will be fundamental for successfully applying adversarial inference to climate, remote sensing, or Earth science tasks. Furthermore, cycle consistency loss defined by the L1 norm does not work well in practice for latent variable translation models - an image mapped to the cloudy domain should be considered cycle-consistent regardless of the spatial position of the cloud, given one or more were generated, as highlighted by the setting of the problem in Section 2.3. 

In this work we have demonstrated the potential of the proposed approach for producing cleaner data for remote sensing applications, and identified some of the weaknesses and shortcomings of DGMs, which are to be addressed in future.


\medskip

\small
\bibliographystyle{plainnat}

\bibliography{neurips_2019}
\end{document}